\documentclass[twocolumn,prb,groupeaddress,showpacs,rapid,floatfix]{revtex4}

\usepackage{graphicx}
\usepackage{bm}

\begin{document}

\title{Magnetism and Superconductivity in Ce$_2$RhIn$_8$}

\author{M.~Nicklas}
\author{V.~A.~Sidorov}
 \altaffiliation{Permanent address: Institute for High Pressure
 Physics, Russian Academy of Sciences, Troitsk, Russia.}
\author{H.~A.~Borges}
 \altaffiliation{Permanent address:  Departamento de F\'{i}sica, Pontif\'{i}cia Universidade Cat\'{o}lica do Rio de Janeiro, Brazil.}
\author{P.~G.~Pagliuso}
\author{C.~Petrovic}
 \altaffiliation{Permanent address: Brookhaven National Laboratory, Upton, NY}
\author{Z.~Fisk}
 \altaffiliation{Permanent address: NHMFL, Florida State University, Tallahassee, FL.}
\author{J.~L.~Sarrao}
\author{J.~D.~Thompson}

\affiliation {Los Alamos National Laboratory, Los Alamos, NM
87545}

\date{\today}

\begin{abstract}
We report the discovery of pressure-induced superconductivity,
with $T_c=2$~K, in the heavy-fermion antiferromagnet
Ce$_2$RhIn$_8$, where superconductivity and magnetic order coexist
over an extended pressure interval. A $T$-linear resistivity in
the normal state, accessed by an applied magnetic field, does not
appear to derive from the existence of a 2D quantum-critical
spin-density wave.
\end{abstract}


\maketitle

Discoveries of pressure-induced superconductivity in several
cerium-based heavy-fermion antiferromagnets have provided a
qualitative perspective on the complex relationship between
magnetism and superonductivity in these highly correlated systems.
\cite{mathur98} CeIn$_3$ is a typical example. Application of
pressure suppresses its N\'{e}el temperature from $T_N\sim$~10 K
at atmospheric pressure to zero temperature at a critical
pressure $P_c\sim$~2.5 GPa.\cite{walker} Neutron-diffraction
studies\cite{morin} show that the ordered 4f-moment decreases
with $T_N(P)$, behavior also reflected in a monotonic depression
of a specific heat anomaly at $T_N$ that disappears as $P$
approaches $P_c$.\cite{knebel} This evolution in magnetic
properties arises from a pressure-induced increase in
hybridization between Ce's $f$-electron and conduction-band
electrons. Near $P_c$, the electrical resistivity assumes a
quasilinear temperature dependence, in contrast to $\rho\propto
T^2$ expected of a Landau Fermi liquid, and is consistent with
quasiparticle scattering from a quantum-critical spin-density
wave.\cite{mathur98} Damped spin fluctuations in this critical
region mediate Cooper pairing in the unconventional
superconducting state that emerges in a narrow pressure range
centered around $P_c$.\cite{mathur98} Like CeIn$_3$, which has a
maximum $T_c\approx0.25$~K, other examples in this class have
$T_c$'s well below 1~K and superconductivity appears only in the
'cleanest' samples with a long electronic mean-free path.

A counter-example to this view is UPd$_2$Al$_3$ in which
local-moment antiferromagnetism and unconventional
superconductivity coexist from atmospheric to high
pressures.\cite{caspary} In this case, the three 5$f$-electrons in
U assume dual characters: two localized $f$'s are responsible for
antiferromagnetism at $T_N$=14.5~K and the other hybridizes with
conduction states to form a liquid of heavy quasiparticles that
becomes unstable below 2~K with respect to a pairing interaction
derived from dispersive excitations of the ordered
moments.\cite{sato}

In both examples, unconventional superconductivity is mediated by
a magnetic interaction, but the bosonic excitations are
distinctly different. In the Ce superconductors there is only a
single $f$ electron participating in both magnetism and
superconductivity through its hybridization with itinerant
electrons; whereas, there is a functional separation of $f$
electrons in UPd$_2$Al$_3$. In the following, we present
pressure-dependent measurements of the heavy-fermion
antiferromagnet Ce$_2$RhIn$_8$ in which magnetic order and
superconductivity appear to coexist over a rather broad pressure
range and are accompanied by an unexpected $T$-linear variation
in electrical resistivity. These results suggest that
Ce$_2$RhIn$_8$ and perhaps the structurally-related compound
CeRhIn$_5$ present a different example of the interplay between
magnetism and superconductivity in strongly correlated matter.

Ce$_2$RhIn$_8$ is a member of the family of heavy-fermion
antiferromagnets Ce$_n$RhIn$_{3n+2}$ composed of $n$ layers of
CeIn$_3$ separated by a single layer of RhIn$_2$, a sequence
repeated along the tetragonal $c$-axis.\cite{grin} The $n=1$
member, CeRhIn$_5$, becomes superconducting at pressures above
1.6 GPa with a $T_c$ exceeding 2~K,\cite{hegger00} nearly an
order of magnitude higher than the infinite layer member CeIn$_3$.
Inserting a second layer of CeIn$_3$ into CeRhIn$_5$ gives
Ce$_2$RhIn$_8$, which orders in a commensurate antiferromagnetic
structure at 2.8~K with an ordered moment of 0.55
$\mu_{B}$,\cite{bao01} slightly reduced by Kondo-spin
compensation from the moment expected in the ground-state
crystal-field doublet. It undergoes a second transition to an
incommensurate magnetic structure at
$T_{LN}$=1.65~K,\cite{malinowski02} which, as will be shown, is
irrelevant to the superconductivity that appears with applied
pressure.

Ce$_2$RhIn$_8$ single crystals were grown out of excess In flux.
X-ray diffraction on powdered crystals revealed single-phase
material in the primitive tetragonal Ho$_2$CoGa$_8$ structure
with lattice parameters $a=0.44665~{\rm nm}$ and $c=1.2244~{\rm
nm}$ at room temperature. There was no evidence for intergrowth of
CeRhIn$_5$. Four-probe $ac$-resistance measurements were made
with current flow in the $(a,b)$-plane. Clamp-type cells
generated hydrostatic pressures to 0.6 GPa for dc magnetization
and 2.3 GPa for resistivity measurements. Flourinert-75 served as
the pressure medium. Hydrostatic pressures to 5.0 GPa were
produced in a toroidal anvil cell using a glycerol-water mixture.
\cite{sidorov91} In both cases, the superconducting transition of
Pb (Sn), which served as a pressure gauge, remained sharp at all
pressures, indicating a pressure gradient of less than 1-2\% of
the applied pressure.

\begin{figure}[t]
\includegraphics[angle=0,width=70mm,clip]{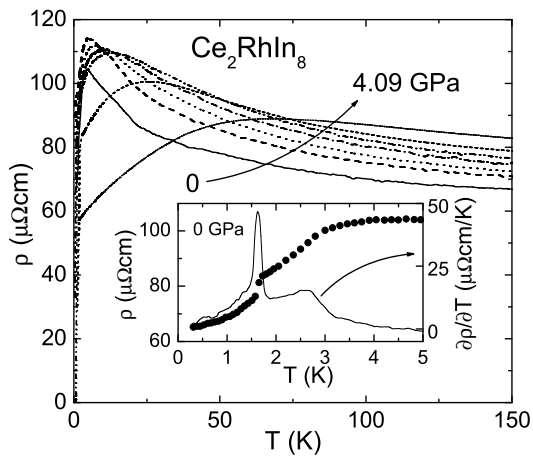}
\caption{\label{fig1} Temperature dependence of the in-plane
resistivity of Ce$_2$RhIn$_8$ at various fixed pressures from
ambient pressure to 4.09~GPa. The inset shows the $P=0$
low-temperature resistivity and its derivative. The magnetic
transitions are clearly indicated at $T_N=2.8$ K and
$T_{LN}=1.65$~ K, respectively. }
\end{figure}

The overall behavior of the resistivity is shown in
fig.~\ref{fig1}. The high temperature resistivity increases with
increasing pressure over the temperature range between about 25~K
and room-temperature. There is well-defined maximum in the
resistivity at $T_{\rm max}=5$~K that initially decreases with
$P$ before increasing at a rate of $\sim$20~K/GPa for
$P\gtrsim2.0$~GPa. This initial negative $\partial T_{\rm
max}/\partial P$ is unexpected for a Ce-based compound but is
found in CeRhIn$_5$.\cite{hegger00,muramatsu01} In that case,
$T_{\rm max}(P)$ followed the pressure dependence of a maximum in
the static susceptibility that is produced by the development of
anisotropic antiferromagnetic correlations above
$T_N$.\cite{baoandaeppli} Presumably, the initial decrease in
$T_{\rm max}(P)$ in Ce$_2$RhIn$_8$ has the same origin as in
CeRhIn$_5$. The increase in $T_{\rm max}(P)$ at higher pressures
is due to a shift of the characteristic spin-fluctuation
temperature to progressively higher energies.

The inset of Fig.~1 shows the low-temperature resistivity and its
derivative at atmospheric pressure. Compared to the typically
small residual resistivity $\rho_0\approx1~{\rm \mu\Omega cm}$ of
CeIn$_3$ and CeRhIn$_5$,\cite{mathur98, hegger00} $\rho_0$ for
Ce$_2$RhIn$_8$ is one to two orders of magnitude higher. A low
residual resistivity ratio and high $\rho_0$ are reproduced in
all of many high-quality crystals of Ce$_2$RhIn$_8$ we have
studied, and, therefore, appears to be an intrinsic property of
this compound. In spite of the high resistivity, $\rho(T)$ and
$\partial\rho/\partial T$ clearly reveal the commensurate and
incommensurate antiferromagnetic transitions at $T_N=2.8$~K and
$T_{LN}=1.65$~K, respectively. Using the data in Fig.~\ref{fig1}
and the maxima in $\partial\rho/\partial T$ to track the pressure
evolution of these phase transitions, we find that $T_N(P)$
decreases linearly at a rate $\partial T_{N}/\partial
P\approx-0.76$~K/GPa. This slope is confirmed by
dc-susceptibility measurements to 0.5~GPa.

\begin{figure}[t]
\includegraphics[angle=0,width=70mm,clip]{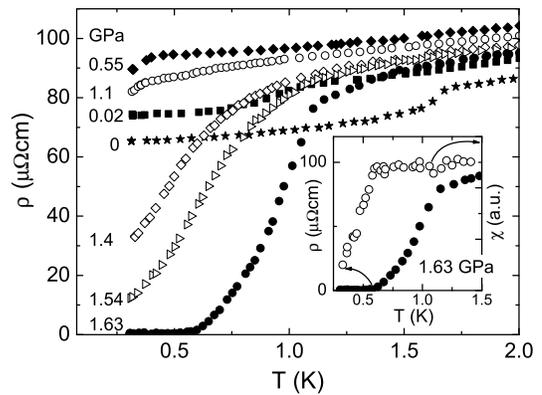}
\caption{\label{fig2} Low-temperature resistivity of the in-plane
resistivity of Ce$_2$RhIn$_8$ at various fixed pressures. See
text for details. With increasing pressure, a zero-resistivity
and diamagnetic state evolves below 600~mK at 1.63~GPa as shown in
the inset.}
\end{figure}

\begin{figure}[b]
\includegraphics[angle=0,width=70mm,clip]{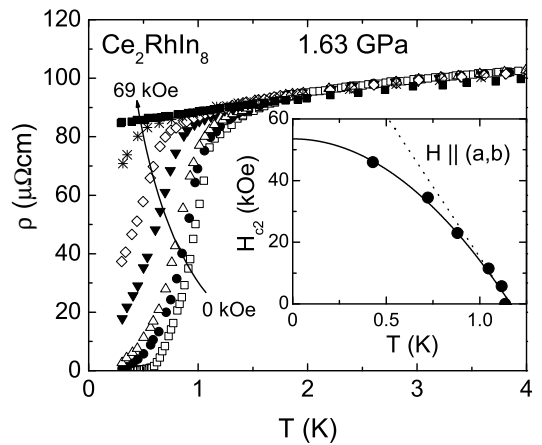}
\caption{\label{fig3} Effect of a magnetic field on the
resistivity of Ce$_2$RhIn$_8$ at $P=1.63$~GPa and in various
magnetic fields up to 69~kOe applied parallel to the basal plane.
The inset shows $H_{c2}$ as a function of the temperature. The
solid curve is a fit described in the text. The initial slope of
$H_{c2}(T)$ is indicated by the dotted line.}
\end{figure}

Figure ~\ref{fig2} gives a detailed view of the low-temperature
resistivity at intermediate pressures. We see that the
incommensurate transition is very sensitive to pressure. $T_{LN}$
shifts from 1.65~K at ambient pressure to 0.95~K at 0.02 GPa.
This gives an estimate of a critical pressure
$P_{c,LN}\approx(0.04\pm0.01)$ GPa for suppressing $T_{LN}$ and a
corresponding slope of $\partial T_{LN}/\partial
P\approx-(43\pm15)$ K/GPa that is consistent with $\partial
T_{LN}/\partial P$ derived from Ehrenfest's relation and
measurements of the low-temperature specific heat and volume
thermal expansion.\cite{malinowski02} Therefore, only the
commensurate phase survives for $P\gtrsim$ 0.04 GPa. At 0.55~GPa
a weak decrease in the resistivity appears at $T_?=420$~mK.
Nearly the same temperature dependence of the resistivity and
same $T_?$ are found for 0.69 and 0.89~GPa (not shown). We do not
know the origin of this feature. At 1.10~GPa the data develop a
steeper slope below $\sim$1~K followed by a kink near
$T_c=380$~mK. The kink shifts continuously to higher temperatures
with increasing pressure and evolves smoothly into a
zero-resistance state below 600~mK at 1.63~GPa. Measurements of
the ac susceptibility, plotted in the inset of Fig.~\ref{fig2},
show the onset of a diamagnetic response at the same temperature
where the resistance goes to zero. Although perfect diamagnetism
could not be observed in the experimentally accessible
temperature range, it is clear from the size of the signal change
that the diamagnetic response is due to bulk superconductivity.
Reproducibility of a zero-resistance state for $P\geq~1.6$~GPa
was confirmed on another crystal.

To provide additional confirmation of bulk superconductivity, we
determined the upper critical field $H_{c2}(T)$ at 1.63 GPa using
data plotted in Fig.~\ref{fig3}. The resistive onset defines
$H_{c2}(T)$ which is shown in the inset. A fit of
$H_{c2}\propto(H_{c2}(T=0)-H_{c2}(T))^2$ describes the data
reasonably well with $H_{c2}(0)=53.6$~kOe and an initial slope
$-{\rm d}H_{c2}/{\rm d}T\mid_{T=Tc} = 91.8$~kOe/K.\cite{Hc2} The
Ginzburg-Landau coherence length in the $c$-axis direction
$\xi_{GL}= [\frac{\Phi_0}{2\pi
H_{c2}(0)}]^{0.5}\approx7.7{\rm~nm}$, which is comparable to the
volume-averaged electronic mean-free path, ${\it
l}\approx6.5{\rm~nm}$.\cite{calc} The dirty-limit relationship
$-dH_{c2}/dT\mid_{T=Tc}\,\propto\rho_0\gamma$ gives
$\gamma\approx0.20{\rm~J/molCe K^2}$ at 1.63~GPa, which is
one-half the value measured directly at atmospheric pressure just
above $T_N$. Halving of the Sommerfeld coefficient at 1.63 GPa is
expected from the relationship $\gamma(P)\propto1/T_{\rm max}(P)$
obeyed by several heavy-fermion systems and our observation that
$T_{\rm max}(1.63)/T_{\rm max}(0)=2.5$ in Ce$_2$RhIn$_8$.
Furthermore, bulk superconductivity evolves out of a distinctly
non-Fermi-liquid-like state. A fit to $\rho(T)$ at 1.63~GPa and
69~kOe, solid squares in Fig.~3, gives $\rho=\rho_0+A'T^n{\rm,
with~} n=0.95\pm0.05$ for $0.3{\rm~K}\leq T\leq 1.8{\rm~K}$. An
approximately $T$-linear resistivity also is found above $T_c$ in
CeRhIn$_5$\cite{muramatsu01} and is qualitatively different from
the $T^{\sim1.6}$ dependence observed in CeIn$_3$ near its
critical pressure.\cite{mathur98}

\begin{figure}[t]
\includegraphics[angle=0,width=70mm,clip]{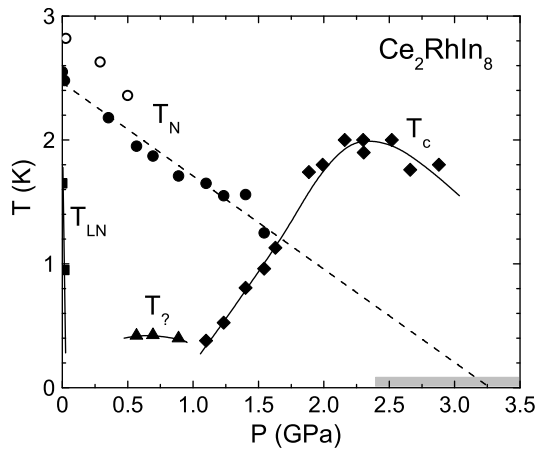}
\caption{\label{fig4} The temperature-pressure phase diagram for
Ce$_2$RhIn$_8$ determined by $\rho(T)$ (solid symbols) and
dc-magnetization (open circles). The lines are guides to the
eyes.}
\end{figure}

Measurements to 5.0~GPa, but for $T\geq1~K$, show the onset of
superconductivity reaching a maximum of 2.0~K near 2.3~GPa before
decreasing below 1~K above 3.5~GPa. See Fig.~\ref{fig4}. $T_c$
reaches a maximum close to the pressure at which $T_N$
extrapolates to zero. Unless $T_N$ drops precipitously above
$\sim$1.5~GPa, superconductivity and local-moment, commensurate
antiferromagnetism coexist over a substantial range of pressures.
A critical pressure of $\sim$2.5~GPa and a 'dome' of
superconductivity with a maximum $T_c$ centered near the
extrapolated critical pressure also are found in
CeIn$_3$;\cite{mathur98} however, the maximum $T_c$ of
Ce$_2$RhIn$_8$ reaches a value comparable to $T_N(P=0)$, as it
does in CeRhIn$_5$,\cite{hegger00, muramatsu01} and is an order
of magnitude higher than in CeIn$_3$, where $T_c/T_N\simeq$~
0.025. Interestingly, the 'dome' of superconductivity exists over
a rather broad pressure range, at least 2.5~GPa, in
Ce$_2$RhIn$_8$ but is very narrow, $\sim$0.4~GPa, in CeIn$_3$;
the pressure range over which superconductivity exists scales
roughly with $T_c$.

To interpret these observations, we first consider the
perspective developed from other examples of $P$-induced
superconductivity in Ce-based systems. For magnetically mediated
superconductivity, $T_c\propto T_{sf}$, where $T_{sf}$ is the
characteristic spin-fluctuation temperature that is inversely
proportional to the specific heat Sommerfeld coefficient
$\gamma$.\cite{nakamura} For CeRhIn$_5$ ,\cite{fisher}
Ce$_2$RhIn$_8$ and CeIn$_3$,\cite{knebel} $\gamma\sim$ 0.4, 0.2
and 0.37 ${\rm~J/mol\,Ce K^2}$, respectively, at $P\lesssim P_c$.
With all other factors equal, $T_c$'s, then, should be
approximately ($\pm50 \%$) the same within this family of
materials; instead, $T_c$'s of the layered compounds are much
higher than in CeIn$_3$. Though $T_{sf}$ sets the overall scale
for the magnitude of $T_c$, $T_c$ also depends on the effective
dimensionality of the spin fluctuations and the electronic
structure: reduced dimensionality favors a higher
$T_c$.\cite{monthoux}

Support for this scenario comes from conventional models of
antiferromagnetic quantum criticality.\cite{millis} These models
predict that, near $P_c$, $T_N\propto (P_c-P)^{\frac{z}{d}}$ and
$\rho(T) \propto T^{\frac{d}{z}}$, where the dynamical exponent
$z$=2 and $d$ is the effective dimensionality of the
spin-fluctuation spectrum. Experimental observations on CeIn$_3$
are consistent with theoretical predictions for\cite{mathur98}
$d$=3 and with $d$=2 in Ce$_2$RhIn$_8$ and provide a plausible
explanation for the unexpectedly high $T_c$ of Ce$_2$RhIn$_8$. In
this picture, the d-wave superconductivity\cite{fisher} and even
somewhat higher $T_c$ of CeRhIn$_5$ would be attributable to more
nearly optimal matching of the momentum dependence of the dynamic
spin susceptibility $\chi(\bf{q},\omega)$ to its quasi-2D
electronic structure.\cite{dHvA} Further, this interpretation
supports superconductivity existing over a much wider range of
pressures in Ce$_2$RhIn$_8$ than in CeIn$_3$ because the
effective pairing interaction is expected to be stronger in
quasi-2D than in 3D.\cite{monthoux}

Though providing a qualitative account of our observations, the
interpretation outlined above relies on a model\cite{millis} in
which the non-Fermi-liquid temperature dependence of $\rho(T)$
arises from Bragg diffraction of heavy quasiparticles off a
quantum-critical spin-density wave (SDW). In this case, the
scattering is critical only on 'hot' portions of the Fermi
surface spanned by the antiferromagnetic ordering wave-vector
{\bf Q}; whereas, other parts of the Fermi surface are
unaffected.\cite{coleman} Unless all of the Fermi surface is hot,
the resistivity should vary as $T^{1+\epsilon}$, where $0<
\epsilon\ <1$, in contradiction to our observations and
those\cite{muramatsu01} on CeRhIn$_5$. Neutron-diffraction
studies of CeRhIn$_5$\cite{llobet} show that {\bf Q} and the
local moment remain well defined and weakly changing as
$P\rightarrow P_c$, supporting speculation of similar behavior in
Ce$_2$RhIn$_8$. We cannot rule out the possibility that most of
the Fermi surface is exactly spanned by {\bf Q} in both
compounds, but this seems very unlikely.

Alternatively, we speculate that the entire Fermi surface is hot
and $\rho(T)$ takes a non-Fermi-liquid form because the quantum
criticality is local.\cite{coleman,si} At a local
quantum-critical point, $\chi(\bf{q},\omega)$ has an anomalous
frequency dependence throughout the entire Brillouin zone and not
just at {\bf Q}. Unlike the weak-coupling SDW limit, local
criticality, which is facilitated by 2-dimensionality, requires
the physics of a Kondo lattice, i.e., a local moment coupled
antiferromagnetically to a bath of itinerant spins and a
fluctuating field produced by surrounding {\it local}
moments.\cite{si} The nature of superconductivity that might
develop near a local quantum-critical point remains to be
investigated, but, because the basic interactions are
antiferromagnetic, as in the SDW limit, superconductivity of
d-wave symmetry would be expected.\cite{si2} Unconventional
superconductivity near a quantum-critical SDW is favored when
$\xi_{GL} \ll {\it l}$,\cite{mathur98} which is satisfied in
CeIn$_3$; however, ${\it l}/\xi_{GL}\sim 1$ in Ce$_2$RhIn$_8$ and
$\sim 3$ in CeRhIn$_5$. This, the much higher $T_c$'s, quasi
2-dimensionality and unexpected $\rho\propto T$ suggest that
unconventional superconductivity in the latter two compounds may
be mediated by qualitatively different spin fluctuations than in
CeIn$_3$. Finally, we note that a theory\cite{zhang} unifying the
order parameters of antiferromagnetism and d-wave
superconductivity predicts a $T-P$ phase diagram like that shown
in Fig. 4 and allows $T_c/T_N\sim 1$ as found in Ce$_2$RhIn$_8$.

We thank Q. Si for helpful discussions and M. Hundley for
screening several Ce$_2$RhIn$_8$ samples. Work at Los Alamos was
performed under the auspices of the US DOE.


\begin{thebibliography}{100}

\bibitem{mathur98} N. D. Mathur, F. M. Grosche, S. R. Julian, I. R. Walker D. M. Freye, R. K. W. Hasselwimmer, and G. G. Lonzarich, Nature {\bf 394}, 39
(1998).

\bibitem{walker} I. R. Walker, F. M. Grosche, D. M. Freye, and G. G. Lonzarich, Physica C {\bf282}, 303 (1997).

\bibitem{morin} P. Morin, C. Vettier, J. Flouquet, M. Konczykowski, Y. Lassailly, J.-M. Mignot, and U. Welp, J. Low Temp. Phys. {\bf70},
377 (1988).

\bibitem{knebel} G. Knebel, D. Braithwaite, P. C. Canfield, G.
Lapertot, and J. Flouquet, High Press. Res. {\bf22}, 167 (2002).

\bibitem{caspary} R. Caspary, P. Hellmann, M. Keller, G. Sparn, C. Wassilew, R. K{\"o}hler, C. Geibel, C. Schank, F. Steglich, and N. E. Phillips, Phys. Rev. Lett.
{\bf71}, 2146 (1993).

\bibitem{sato} N. K. Sato, N. Aso, K. Miyake, R. Shiina, P. Thalmeier, G.
Varelogiannis, C. Geibel, F. Steglich, P. Fulde, and T.
Komatsubara, Nature {\bf410}, 340 (2001).

\bibitem{grin} Yu. N. Grin', Ya. P. Yarmolyuk, and E. I.  Gladyshevskii,  Sov. Phys. Crystallogr.
{\bf 24}, 137 (1979).

\bibitem{hegger00}H. Hegger, C. Petrovic, E. G. Moshopoulou, M. F. Hundley, J. L. Sarrao, Z. Fisk, and J. D. Thompson, Phys. Rev. Lett. {\bf 84},
4986 (2000).

\bibitem{bao01} W. Bao, P. G. Pagliuso, J. L. Sarrao, J. D. Thompson, and Z. Fisk, Phys. Rev. B {\bf
64}, 020401(R) (2001).

\bibitem{malinowski02} A. Malinowski, M. F. Hundley, P. G. Pagliuso, J. L. Sarrao, N. O. Moreno, M. Nicklas, and J. D. Thompson, unpublished; W. Bao, unpublished.

\bibitem{sidorov91} V. A. Sidorov and O. B. Tsiok, Fizika i Tekhnika Vysokikh Davlenii {\bf 1}, 74 (1991).

\bibitem{muramatsu01} T. Muramatsu, N. Tateiwa, T. C. Kobayashi, K. Shimizu, K. Amaya, D. Aoki, H. Shishido, Y. Haga, and
Y. \={O}nuki, J. Phys. Soc. Jpn. {\bf 70}, 3362 (2001).

\bibitem{baoandaeppli} W. Bao, G. Aeppli, J. W. Lynn, P. G. Pagliuso, J. L. Sarrao, M. F. Hundley, J. D. Thompson, and Z. Fisk, Phys. Rev. B {\bf
65}, 100505 (2002).

\bibitem{Hc2} Similar measurements at 15.4~kbar give $H_{c2}(0)=44.8$ kOe and $-{\rm d}H_{c2}/{\rm
d}T\mid_{T=T_c}=93.4$~kOe/K.

\bibitem{calc} The mean free path was estimated by assuming $\xi_{GL}\approx \xi_{BCS}$ and by following relationships given in T. P.
Orlando, E.J. McNiff, S. Foner, and M. R. Beasley Phys. Rev. B
{\bf 19}, 4545 (1979).

\bibitem{nakamura} See, for example, S. Nakamura, T. Moriya and K.
Ueda, J. Phys. Soc. Jpn. {\bf 65}, 4026 (1996).

\bibitem{fisher} R. A. Fisher, F. Bouquet, N. E. Phillips, M. F. Hundley, P. G. Pagliuso, J. L. Sarrao, Z. Fisk, and J. D. Thompson, Phys. Rev. B {\bf 65}, 224509 (2002).

\bibitem{monthoux}P. Monthoux and G. G. Lonzarich, Phys. Rev. B
{\bf63}, 054529 (2001); Phys. Rev. B {\bf59}, 14598 (1999);
cond-mat/0207556.

\bibitem{millis} J. A. Hertz, Phys. Rev. B {\bf 14}, 1165 (1976);
A. J. Millis Phys. Rev. B {\bf 48}, 7183 (1993); U. Z\"{u}licke
and A. J. Millis Phys. Rev. B {\bf 51}, 8996
(1995).

\bibitem{dHvA} D. Hall, E. C. Palm, T. P. Murphy, S. W. Tozer, C. Petrovic, E. Miller-Ricci, L. Peabody,
C. Q. H. Li, U. Alver, R. G. Goodrich, J. L. Sarrao, P. G.
Pagliuso, J. M. Wills, and Z. Fisk, Phys. Rev. B {\bf 64}, 064506
(2001).

\bibitem{coleman} See, for example, P. Coleman, C P\'{e}pin, Q. Si, and R. Ramazashvili, J. Phys. Condens. Matter {\bf 13}, R723 (2001);
P. Coleman and C. Pepin, Physica B {\bf312}, 397 (2002).

\bibitem{llobet} A. Llobet {\it et al.}, unpublished; S. Majumdar, G. Balakrishnan, M. R. Lees, D. McK Paul, and G. J. McIntyre, cond-mat/0205667.

\bibitem{si} Q. Si, S. Rabello, K. Ingersent, and J. L. Smith, Nature {\bf413}, 804 (2001).
Conditions for local quantum criticality appear to be satisfied
in CeRhIn$_5$ as $P\rightarrow P_c$ but remain to be established
for Ce$_2$RhIn$_8$ under pressure.

\bibitem{si2} Q. Si, unpublished.

\bibitem{zhang} S. C. Zhang, Science {\bf275}, 1089 (1997);  A.
Dorneich, E. Arrigoni, M. J\"{o}stingmeier, W. Hanke, and S.-C.
Zhang, cond-mat/0207528.

\end{thebibliography}
\end{document}